\documentclass[amsmath,amssymb,reprint,twocolumn,superscriptaddress, prb,nobibnotes]{revtex4-1}

\usepackage[utf8]{inputenc}
\usepackage[T1]{fontenc}

\usepackage[english]{babel}

\usepackage{amsmath,amssymb,amsfonts}
\usepackage{bbm,mathtools}
\usepackage{amstext, mathrsfs, textcomp}
\usepackage{mathptmx}
\usepackage{bigints}  
\usepackage{nicefrac}
\usepackage{multirow}
\usepackage{dcolumn}
\usepackage{bm}

\usepackage{subfigure}
\usepackage{graphicx}
\usepackage{xcolor}
\usepackage[export]{adjustbox}

\usepackage[colorlinks=true, citecolor=blue, urlcolor=blue, linkcolor=blue]{hyperref}

\usepackage{changes}

\graphicspath{{Figures/}{TOC_ACS/}}

\newcommand{\TITLE}{Deep learning enabled strategies for modelling of complex aperiodic plasmonic metasurfaces of arbitrary size}

\newcommand{\manualref}[1]{#1}  

\begin{document}
	\title{\TITLE}
	
	\author{\firstname{Clément} \surname{Majorel}}
	\affiliation{CEMES-CNRS, Universit\'e de Toulouse, CNRS, UPS, Toulouse, France}
	
	\author{\firstname{Christian} \surname{Girard}}
	\affiliation{CEMES-CNRS, Universit\'e de Toulouse, CNRS, UPS, Toulouse, France}
	
	\author{\firstname{Arnaud} \surname{Arbouet}}
	\affiliation{CEMES-CNRS, Universit\'e de Toulouse, CNRS, UPS, Toulouse, France}
	
	\author{\firstname{Otto L.} \surname{Muskens}}
	\affiliation{Physics and Astronomy, Faculty of Engineering and Physical Sciences, University of Southampton, Southampton, UK}
	
	\author{\firstname{Peter R.} \surname{Wiecha}}
	\email[e-mail~: ]{pwiecha@laas.fr}
	\affiliation{LAAS, Universit\'e de Toulouse, CNRS, Toulouse, France}

	\begin{abstract}
		\section*{Abstract}
		Optical interactions have an important impact on the optical response of nanostructures in complex environments. 
		Accounting for interactions in large ensembles of structures requires computationally demanding numerical calculations.
		In particular if no periodicity can be exploited, full field simulations can become prohibitively expensive. 
		Here we propose a method for the numerical description of aperiodic assemblies of plasmonic nanostructures. Our approach is based on dressed polarizabilities, which are conventionally very expensive to calculate, a problem which we alleviate using a deep convolutional neural network as surrogate model.
		We demonstrate that the method offers high accuracy with errors in the order of a percent.
		In cases where the interactions are predominantly short-range, e.g. for out-of-plane illumination of planar metasurfaces, it can be used to describe aperiodic metasurfaces of basically unlimited size, containing many thousands of unordered plasmonic nanostructures. 
		We furthermore show that the model is capable to spectrally resolve coupling effects.
		The approach is therefore of highest interest for the field of metasurfaces. It provides significant advantages in applications like homogenization of large aperiodic planar metastructures or the design of sophisticated wavefronts at the micrometer scale, where optical interactions play a crucial role.
		
		\textbf{Keywords:} Deep learning, nanophotonics, rapid nano-optics simulations, plasmonic metasurfaces, non-periodic nano-structures
	\end{abstract}

	\maketitle

	\section*{Introduction}
	
	Optical metasurfaces, arrays of nanostructures or other assemblies of many photonic particles have enabled countless exciting applications in the past few decades.\cite{genevetRecentAdvancesPlanar2017, kravetsPlasmonicSurfaceLattice2018}
	From a fabrication point of view, perfectly controlled, highly ordered distributions of nanostructures require usually sophisticated fabrication processes like electron-beam lithography (EBL), which are very expensive and do not scale well to large areas and high throughput. 
	Mass-fabrication compatible techniques on the other hand are often based on chemical synthesis and usually favorable for creating non-periodic distributions of nanostructures.\cite{luChemicalSynthesisNovel2009, siddiqueScalableControlledSelfassembly2017, marcoBroadbandForwardLight2021}
	But disorder in meta-structures can also be introduced on purpose. 
	For instance, hyperuniform aperiodic structures can be specifically designed to render photonic devices robust against environmental influences. \cite{milosevicHyperuniformDisorderedWaveguides2019}
	In a similar approach, randomness can be for instance used to render metasurface designs polarization independent. \cite{dupreDesignRandomMetasurface2018}
	Aperiodic, yet ordered metasurfaces have also been used for near-field shaping, for which optical interactions between plasmonic elements usually play a crucial role.\cite{miscuglioPlanarAperiodicArrays2019}
	Disorder can also play a role in the collective response of very large collections of nanostructures.\cite{solisUltimateNanoplasmonicsModeling2014, rahimzadeganDisorderInducedPhaseTransitions2019}
	For example, recent studies investigating the transition from homogeneous media to heterogeneous, random assemblies of larger structures revealed unexpected physical effects.\cite{werdehausenModelingOpticalMaterials2020}
	Moreover, designer metasurfaces with multiplexed functionalities like polarization diversity or achromatic responses typically consist of deterministic but complex arrangements of particles. \cite{yulevichOpticalModeControl2015, vekslerMultipleWavefrontShaping2015, wuVersatilePolarizationGeneration2017, wangBroadbandAchromaticOptical2017, maguidDisorderinducedOpticalTransition2017}

	Despite their practical importance, large aperiodic arrangements of photonic nanostructures pose a challenging problem for computational methods.
	Their numerical description requires techniques capable to fully describe optical interactions between non-homogeneously arranged nanostructures, which becomes increasingly difficult for growing sizes of the systems under study.
	This is why in the numerical description of metasurfaces, near-field interactions are often quite radically approximated.
	Optical coupling is often either simply neglected, approximated by an element-wise periodicity during the calculation, or by performing a local phase correction using the phase deviations obtained from consecutive nearest-neighbor calculations.\cite{hsuLocalPhaseMethod2017, patouxChallengesNanofabricationEfficient2021}
	Recently a deep learning based method has been demonstrated to include optical coupling in inverse design for periodic metasurfaces with relatively simple unit cells.\cite{anDeepConvolutionalNeural2021}

	In random arrangements however, the situation is even more complicated. For increasingly large systems of aperiodic nano-scatterers full field simulations can become virtually impossible. 
	Taking interaction effects like near-field coupling or multi-scattering fully into account is possible to a certain extent for instance via T-matrix approaches \cite{egelCELESCUDAacceleratedSimulation2017, devriesPointScatterersClassical1998} or in a dipole approximation with effective polarizabilities \cite{devriesPointScatterersClassical1998, cazeStrongCouplingTwoDimensional2013, leseurProbingTwodimensionalAnderson2014}. 
	In such approach, each constituent of the metasurface is numerically described by a single effective dipole or substituted by a spherical structure \cite{sersicMagnetoelectricPointScattering2011, bowenUsingDiscreteDipole2012, arangoPolarizabilityTensorRetrieval2013, patouxPolarizabilitiesComplexIndividual2020, bernalarangoUnderpinningHybridizationIntuition2014, bertrandGlobalPolarizabilityMatrix2019, munDescribingMetaAtomsUsing2020}.
	In the case of the effective polarizability approximation, an optically interacting ensemble of nanostructures can be mathematically formulated as a system of coupled equations, taking optical interactions between particles rigorously into account. \cite{bowenUsingDiscreteDipole2012}
	However, this approach reaches its limits if the size of the coupled system of dipoles grows exceedingly but also if the interacting particles are too close so that their 3D shape can not be approximated by a point scatterer model.
	
	In the recent past, data-based approaches and in particular deep learning (DL), have been found to provide powerful numerical tools, capable to overcome limitations such as the above described computational cost of complex numerical simulations. 
	DL has been successfully applied to various problems in nanophotonics. 
	In nano-optics and photonics, DL has enabled to efficiently tackle complex problems, such as real time deconstruction of complex optical signals under sparse sampling conditions \cite{kamilovLearningApproachOptical2015, borhaniLearningSeeMultimode2018, rivensonPhaseRecoveryHolographic2018, kurumDeepLearningEnabled2019}, the ability to provide huge accelerations for physics simulations \cite{raissiPhysicsinformedNeuralNetworks2019, wiechaDeepLearningMeets2020, jiangDeepNeuralNetworks2021, blanchard-dionneTeachingOpticsMachine2020} and to perform ultra fast inverse design of photonic nanostructures and metasurface unit cells \cite{liuTrainingDeepNeural2018, liuGenerativeModelInverse2018, jiangFreeFormDiffractiveMetagrating2019, dinsdaleDeepLearningEnabled2021, wiechaDeepLearningNanophotonics2021}.
	
	Here we apply deep learning to predict the optical response of complicated, aperiodic assemblies of plasmonic nanostructures. 
	To this end we employ a dipolar approximation based on dressed polarizabilities. 
	The dressed polarizability concept was used relatively rarely in the past, mainly because the calculation of dressed polarizabilities in complex environments is similary expensive as the full-field simulations themselves. 
	Furthermore, dressed polarizabilities include the response of the environment, hence these are valid only in a fixed geometric configuration and need to be re-calculated even for small changes in the surrounding scene.
	On the other hand, once calculated, dressed polarizabilities take into account all interactions with the local environment, such as with neighbor nanostructures or with interfaces like a substrate, and allow to calculate at practically no cost the response of the particles upon illumination with any polarization.\cite{patouxPolarizabilitiesComplexIndividual2020}
	We demonstrate here that, using a deep learning framework, the result of the expensive calculation can be accurately approximated in many orders of magnitude shorter computation time.
	We show that within a local neighborhood approximation, the method allows to calculate fully non-periodic, disordered meta-structures of arbitrary size.
	Finally we demonstrate that the model can be efficiently trained also on wavelength dependent data, rendering it very useful for the investigation of coupling phenomena in complex nanostructures or for the homogenization of aperiodic, disordered metasurfaces.

	\begin{figure}[t]
		\centering
		\includegraphics[width=\columnwidth]{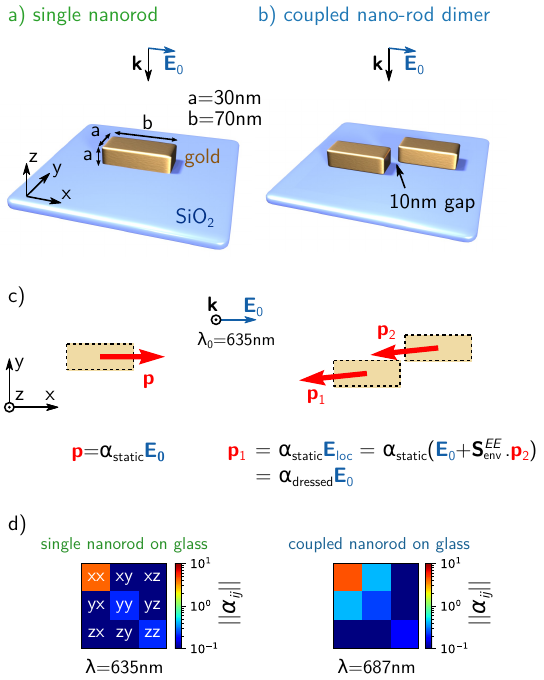}
		\caption{
			Dressed polarizabilities: Illustration of the impact of the local environment on the induced electric dipole moment in a gold nanorod.
			(a) Sketch of a single gold nanorod of size $70\times 30\times 30\,$nm$^3$ lying on a dielectric substrate of refractive index $n_{\text{subst}}=1.45$. 
			(b) Sketch of two gold nanorods on a glass substrate, almost touching at their corners, separated by a gap of $10\,$nm.
			(c) electric dipole moment (real part), induced in the gold nanorods shown on (a-b). The presence of the second rod at a short distance has a severe impact on the optical response due to the field scattered by the neighbor nanorod. The modification of the response can be described by a dressed polarizability tensor $\alpha_{\text{dressed}}$.
			(d) Norm of the dressed polarizability components at the respective resonance wavelength of $\text{Im}(\alpha_{xx})$.
		}
		\label{fig:polarization_sketch}
	\end{figure}

	\section*{Methods}
	
	\subsection*{Dressed polarizabilities}
	
	In a dipolar approximation the polarizability $\alpha$ describes the optical response of a nanostructure and relates the electric polarization $\mathbf{p}$ of an effective dipole to the local electric field $\mathbf{E}_{\text{loc}}$ :
	
	\begin{equation}\label{eq:polarizability}
		\mathbf{p} (\omega_0) = \alpha (\omega_0) \cdot \mathbf{E}_{\text{loc}}(\omega_0)\, .
	\end{equation}
	
	For a spherical particle of radius $r_{\text{sphere}}$, very small compared to the wavelength, the polarizability can be described by the permittivity $\epsilon_{\text{sphere}}$ of the material \cite{markelIntroductionMaxwellGarnett2016}:
	
	\begin{equation}\label{eq:polarizability_sphere}
		\alpha (\omega_0) = \frac{\epsilon_{\text{sphere}}(\omega_0)-1}{\epsilon_{\text{sphere}}(\omega_0)+2}\ r_{\text{sphere}}^3 \, .
	\end{equation}
	This dipolar electric polarizability is exact only in the quasistatic regime, but an effective dipolar polarizability can nevertheless be an excellent approximation for nanostructures of intermediate sizes (but still smaller than the wavelength). 
	This is true in particular for far-field observables, because in contrast to higher-order contributions, the dipolar mode couples very efficiently to far-field radiation and hence often dominates the scattering and extinction.
	Even if the shape of a nanostructure is non-isotropic, it is still possible to use an effective polarizability to describe the structure's interaction with an incident electromagnetic wave. 
	In the case of complex nanostructure geometries the polarizability takes the form of a rank two tensor, which, in the limits of a dipolar description of the optical response, can also include polarization conversion effects through its off-diagonal elements.\cite{wiechaPolarizationConversionPlasmonic2017}
	Depending on the nanostructure shape, the effective polarizability can be extracted using analytical or numerical methods.\cite{morozDepolarizationFieldSpheroidal2009, arangoPolarizabilityTensorRetrieval2013, patouxPolarizabilitiesComplexIndividual2020}
	It is worth mentioning at this point, that we consider here only the electric dipolar response.
	Larger dielectric nanostructures can induce magnetic dipole resonances due to optical vortices which are a result of the phase delay of the illumination along the structure.\cite{sersicMagnetoelectricPointScattering2011, kuznetsovMagneticLight2012} Since these magnetic effects in dielectrics are a result of retardation, they are $\mathbf{k}$-vector dependent and therefore in general more difficult to treat in an effective polarizability approximation.\cite{patouxPolarizabilitiesComplexIndividual2020}
	
	The appeal of the effective polarizability approximation is the fact that it is possible to calculate at any location $\mathbf{r}$ the electromagnetic field, scattered by a dipolar particle at position  $\mathbf{r}'$. This is done using the electric-electric Green's tensor $\mathbf{S}^{EE}_{\text{env}}$ of the bare environment, also called field susceptibility:\cite{girardFieldsNanostructures2005}
	\begin{equation}\label{eq:repropa_E}
		\mathbf{E}(\mathbf{r}, \omega_0) = \mathbf{E}_0(\mathbf{r}, \omega_0) + \mathbf{S}^{EE}_{\text{env}}(\mathbf{r},\mathbf{r}',\omega_{0}) \cdot \mathbf{p}(\mathbf{r}', \omega_0) \, .
	\end{equation}
	Likewise, the magnetic field emitted by the electric dipole can be obtained at any location in space via the mixed magnetic-electric field susceptibility $\mathbf{S}^{HE}_{\text{env}}$: \cite{girardOpticalMagneticNearfield1997}
	\begin{equation}\label{eq:repropa_H}
		\mathbf{H}(\mathbf{r}, \omega_0) = \mathbf{H}_0(\mathbf{r}, \omega_0) + \mathbf{S}^{HE}_{\text{env}}(\mathbf{r},\mathbf{r}',\omega_{0}) \cdot \mathbf{p}(\mathbf{r}', \omega_0) \, .
	\end{equation}
	Please note that we assume here a purely electric dipole response of the nanostructure, as will be also further discussed below.
	
	In case of an isolated scatterer, the local field $\mathbf{E}_{\text{loc}}$ at the nanostructure is identical to the illumination field $\mathbf{E}_0$. This case is depicted in figure~\ref{fig:polarization_sketch}a, by the example of a single gold nanorod.
	In complex environments however, the local field can be more or less strongly perturbed -- for instance by interfaces, e.g. of a substrate or due to neighbor nanostructures. The reflection or scattering of these perturbing elements superposes with the illumination:
	\begin{equation}
		\mathbf{E}_{\text{loc}} (\mathbf{r}, \omega_0) = \mathbf{E}_0 (\mathbf{r}, \omega_0) + \mathbf{E}_{\text{perturbations}} (\mathbf{r}, \omega_0) \, .
	\end{equation}
	A simple example is shown in figure~\ref{fig:polarization_sketch}b, where a gold nanorod is laying on a glass substrate, very close to a second rod. 
	Figure~\ref{fig:polarization_sketch}c illustrates with a red arrow the effective dipole moment induced in the gold-rods.
	For this example a plane wave at the resonance wavelength $\lambda_0=635\,$nm of an isolated nanorod is normally incident with a polarization along the rod axis. 
	We see that the effective dipole moments not only flip sign with respect to an isolated gold rod, their diagonal geometric alignment furthermore introduces a component of the electric polarization normal to the field vector of the illumination. 
	Also the resonance wavelength is modified by the interaction.
	The non-zero tensor components and their respective strengths are illustrated by $3\times 3$ color plots  in Fig.~\ref{fig:polarization_sketch}d, shown at each configuration's resonance wavelength.
	
	In complicated environments, the local field and its interaction with a nano-scatterer is in general not trivially known. 
	The determination requires sophisticated numerical simulations. 
	It can be obtained for instance via the Green's Dyadic Method (GDM), which calculates the interactions of all nanostructures using the field susceptibilities associated with the empty environment, followed by a volume discretization of the ensemble of scatterers.\cite{smunevRectangularDipolesDiscrete2015, patouxPolarizabilitiesComplexIndividual2020}
	In the presence of many neighbor nanostructures, full field numerical simulations such as the GDM can be very slow and a dipolar model such as the above introduced effective polarizability approximation appears to be an interesting approach.
	In fact it is possible to define a so-called dressed polarizability tensor $\alpha_{\text{dressed}}$, by including the contribution  $\mathbf{E}_{\text{perturbations}}$ of the environment to the local field into the polarizability tensor of each nano-scatterer. 
	This requires to solve the electromagnetic fields at the location of the nanostructure for arbitrary illuminations such that the environment contribution to the local field can be separated from the illumination. 
	The electric polarization of the effective dipole can in consequence be written as function of the unperturbed incident field:
	\begin{equation}\label{eq:dressed_polarizability}
		\begin{aligned}
			\mathbf{p} (\mathbf{r}_0, \omega_0) 
				& = \alpha (\mathbf{r}_0, \omega_0) \cdot \mathbf{E}_{\text{loc}}(\mathbf{r}_0, \omega_0) \\ 
				& = \alpha_{\text{dressed}} (\mathbf{r}_0, \omega_0) \cdot \mathbf{E}_0(\mathbf{r}_0, \omega_0)\, .
		\end{aligned}
	\end{equation}

	We see that once the dressed polarizabilities $\alpha_{\text{dressed}}$ of all nanostructures in a complex arrangement are known, the result of the light-matter interaction is obtained instantaneously for arbitrary illumination polarizations, by a simple vector product of the unperturbed incident field $\mathbf{E}_0$ with the dressed polarizability tensors of the nanostructures.
	
	In the context of the GDM, the dressed polarizability of a structure can be extracted through a volume discretization, using the concept of a generalized field propagator:\cite{martinGeneralizedFieldPropagator1995}
	\begin{equation}\label{eq:alpha_dressed}
		\begin{multlined}
		\alpha_{\text{dressed}}(\mathbf{r}_0, \omega_{0}) = \\
		v_{\text{cell}}^{2}(\omega_{0})\ 
		\sum_{i}^{\text{NS at }\mathbf{r}_0}\ \
		\sum_{j}^{\text{ensemble}}\ 
		\boldsymbol\chi(\mathbf{r}_{j}, \omega_{0})\, \mathbf{K}^{EE}(\mathbf{r}_{i},\mathbf{r}_{j},\omega_{0})
		\end{multlined}
	\end{equation}
	where the sum over the index $j$ runs over the entire ensemble of nanostructures, thus including the full local ``neighborhood''. 
	The sum over index $i$ (``NS'') includes only the meshpoints of the dressed nanostructure, centered at $\mathbf{r}_0$, of which we wish to calculate the dressed polarizability.
	In addition, $v_{\text{cell}}$ is the volume of the GDM discretization cells and $\boldsymbol\chi$ is the electric susceptibility tensor of the material. 
	$\mathbf{K}^{EE}$ is the generalized field propagator describing electric-electric field coupling in the nanostructure \cite{martinGeneralizedFieldPropagator1995}:
	
	\begin{equation}\label{eq:KEE}
			\mathbf{K}^{EE}(\mathbf{r},\mathbf{r}',\omega_{0}) = \delta(\mathbf{r}-\mathbf{r}') \mathbf{I} + \boldsymbol\chi(\mathbf{r}', \omega_{0})\cdot \mathbf{S}^{EE}(\mathbf{r},\mathbf{r}',\omega_{0})\, ,
	\end{equation}
	where $\delta$ is the Kronecker-delta, $\mathbf{I}$ is the unitary tensor and $\mathbf{S}^{EE}$ is the field susceptibility of the complex environment including all nanostructures.
	The latter can be numerically obtained through a volume discretization of the ensemble of all nanostructures. It hence implicitly contains the self-consistent fields from the Lippmann-Schwinger equation. The interactions with a substrate can be included using according Green's tensors.\cite{paulusAccurateEfficientComputation2000, girardFieldsNanostructures2005, patouxPolarizabilitiesComplexIndividual2020}
	Specifically we use our homemade python toolkit ``pyGDM'' for the extraction of the dressed polarizabilities.\cite{wiechaPyGDMPythonToolkit2018, wiechaPyGDMNewFunctionalities2022}
	We note that since we fully discretize the volume of the nanostructures for the extraction step, the resulting dressed polarizabilities correctly include coupling effects at the level of the true nanostructure surfaces, even though the final result is eventually reduced to a dipolar approximation. Especially in dense configurations it therefore goes significantly beyond a coupled dipole description where, from the beginning, each nanostructure would be represented by a single dipole.
	Here on the other hand, we apply the reduction to a point dipole as the very last step, after having solved the self-consistent fields within the full discretization.
	
	We would like to note two important limitations of the approach here. 
	First, our dressed polarizability describes only the dipolar electric response of a nanostructure, it therefore only works for plasmonic nanostructures or for non-resonant dielectric particles in the long wavelength limit. 
	Resonant dielectric structures induce magnetic effects as a result of the accumulated phase of the incident field. \cite{kuznetsovOpticallyResonantDielectric2016}
	With the phase implicitly being set constant in equation~(\ref{eq:alpha_dressed}), these magnetic effects are neglected in the approximation (see also Ref.~\onlinecite{patouxPolarizabilitiesComplexIndividual2020}).
	The constant phase is also the reason for the second limitation: Since we assume a fixed phase relation in the entire ``neighborhood'', the resulting polarizabilities are also dressed with respect to the angle of incidence and equation~(\ref{eq:alpha_dressed}) implicitly assumes normal incidence. 
	Under these conditions however, all multiple scattering events as well as near-field coupling are included in the dressed polarizability. 
	In consequence, the accuracy of the approximation is excellent, of the order of one percent.
	
	We note that in analogy to the formalism developed in Ref.~\onlinecite{patouxPolarizabilitiesComplexIndividual2020}, a local phase term could be included in equation~(\ref{eq:alpha_dressed}), which would allow to describe non-normal illumination angles, arbitrary wavefronts and also magnetic optical effects.
	Here however, we will limit the demonstration to normal incicence and an electric dipolar response, which by itself reflects a quite widespread configuration.

	\subsection*{Deep learning for the rapid extraction of dressed polarizabilities}

	As discussed above, the extraction of the $\alpha_{\text{dressed}}$ requires a similar computational effort as a full-field simulation, but comes at the price of a reduced accuracy due to the dipolar approximation for each nanostructure. 
	Moreover, in contrast to conventional polarizabilities, their dressed counterparts are valid only for a single particular arrangement of nanostructures.
	A small change in the distribution of scatterers changes the dressed polarizabilities of \textit{all the nanostructures} in the arrangement. 
	Therefore, dressed polarizabilities have been only occasionally used, for instance in very simple environments\cite{castanieAbsorptionOpticalDipole2012} or to determine effective medium permittivities of periodic layers of nanostructures.\cite{wijersOpticsEmbeddedSemiconductor2006, yooEffectivePermittivityResonant2012} 
	In the latter scenario the periodicity is naturally imposed also on the local field and hence each element has the same $\alpha_{\text{dressed}}$ tensor.
	In consequence the computational effort is limited to a single constituent and the dressed polarizability can then be used to extract an effective index for the periodic layer via effective medium theories like the Maxwell-Garnett approximation.\cite{levyMaxwellGarnettTheory1997, markelIntroductionMaxwellGarnett2016, vynckLightCorrelatedDisordered2021}

	In large, aperiodic assemblies of nanostructures on the other hand, the concept of dressed polarizabilities seemed less useful so far, mainly because of the considerable computational effort.
	To overcome this limitation, we present here a data-driven approach based on deep learning, allowing to approximate dressed polarizabilities of many nanostructures in complex assemblies at almost no computational cost. 
	The data-driven approximation is several orders of magnitude faster compared to the conventional calculation of dressed polarizabilities via full field electrodynamics simulations. 
	In combination with the assumption of predominantly short-range interactions, the neural network predictor allows to obtain near- and far-fields for disordered metasurfaces of essentially arbitrary size.

	\begin{figure}[t]
		\centering
		\includegraphics[width=\columnwidth]{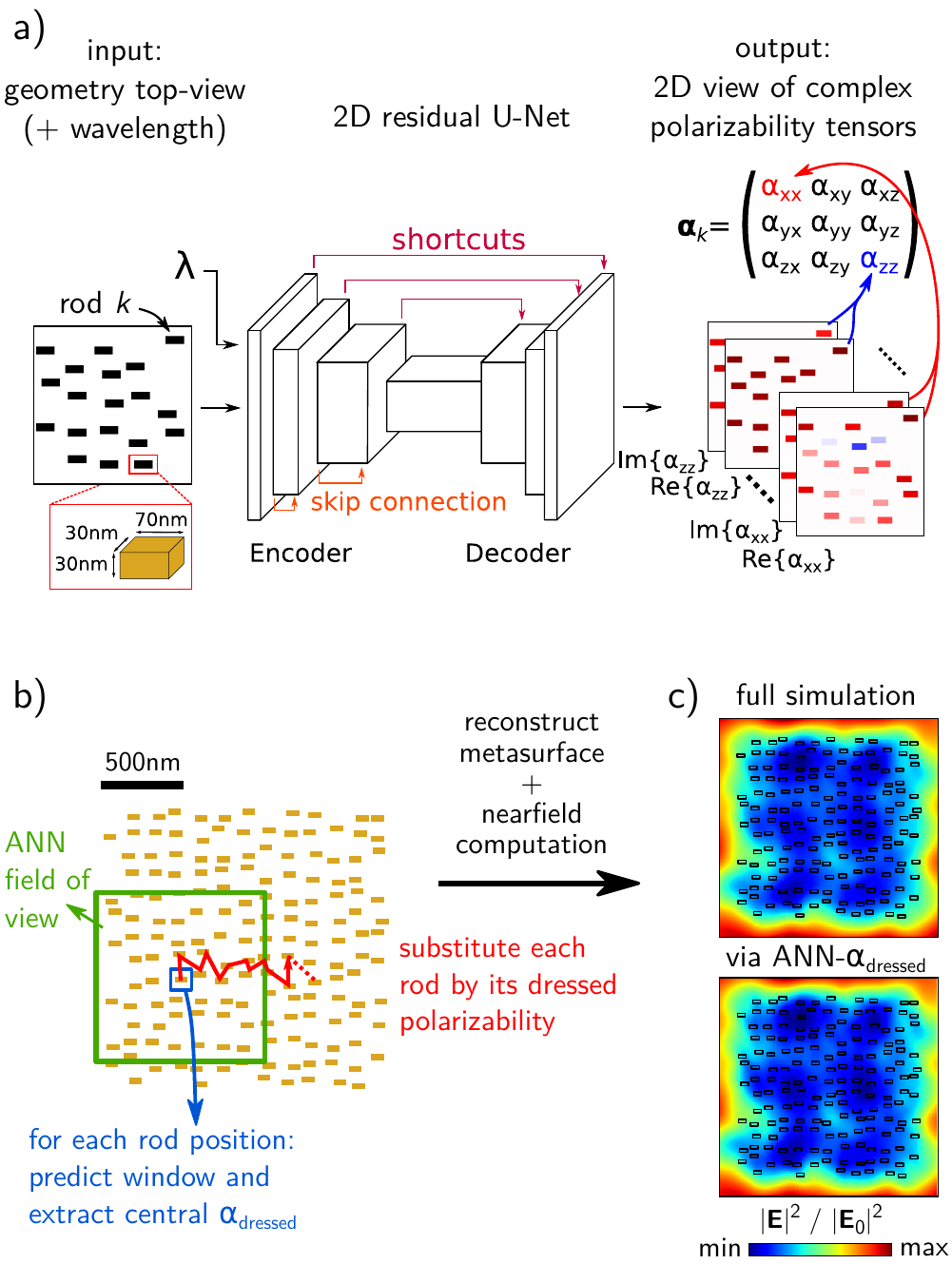}
		\caption{
			a) Sketch of the neural network model.
			The network is based on a symmetric 2D U-Net, built from residual blocks of 4 convolutional layers each.
			The encoder takes as input a 2D top-view of the nanorod arrangement (left), each pixel of the input image corresponds to a step of $10\,$nm. 
			In a second input channel, the evaluation wavelength is fed into the network, expanded to a layer of the same dimension as the geometry input, filled with the wavelength in units of \textmu m.
			The output of the network is the 2D top-view of dressed polarizabilities for each structure (right). Since each polarizability is described by a complex-valued $3\times 3$ tensor, the network output consists in 18 channels, 9 for the real, 9 for the imaginary parts of each tensor component.
			b) Illustration of the local neighborhood approximation procedure. An area of the size of the ANN's input field is centered on each rod in a sequential evaluation procedure, which we refer to as ``dressed polarizability substitution''. From each neural network's prediction, only the dressed polarizability of the central rod is retained.
			c) Qualitative test of the sequential method by the example of a near-field intensity map, $150\,$nm above the substrate under normal illumination at $\lambda_0=635\,$nm. 
			The top panel shows a full simulation, the bottom panel is based on the neural network prediction with a sequential evaluation of each rod in the local approximation. 
		}
		\label{fig:neural_network_model}
	\end{figure}

	\begin{figure*}[t]
		\centering
		\includegraphics[width=\linewidth]{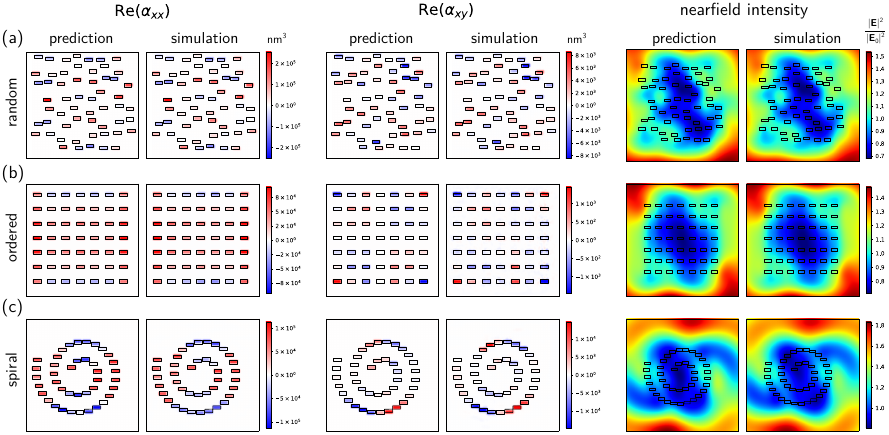}
		\caption{Dressed polarizability predictions compared to simulations.
			a) Random distribution of rods as used for the training and test data-sets.
			b) perfectly ordered 2D array of rods.
			c) spiral as example of an ordered, non-periodic arrangement of rods.
			The colormaps of the polarizability components $\alpha_{xx}$ (two left columns) and $\alpha_{xy}$ (two center columns) show areas of $1\times 1\,$\textmu m$^2$. The near-field maps (two right columns) are calculated $150\,$nm above the substrate interface, under plane wave illumination at $\lambda_0=635\,$nm with left circular polarization, incident from below. The near-field intensity maps show areas of $1.3\times 1.3\,$\textmu m$^2$.
		}
		\label{fig:prediction_examples_fixed_wl}
	\end{figure*}

	\paragraph*{General concept and geometric model}
	Our concept consists in using the complex arrangement of many nanostructures in the plane of the metasurface and to predict approximate dressed polarizabilities simultaneously for all constituents using a deep convolutional neural network (CNN).
	The principle is illustrated in figure~\ref{fig:neural_network_model}a.
	The input area of the neural network -- its \textit{field of view} -- has a size of $1\times 1$\,\textmu m$^2$. 
	Our goal is to use this neural network then to evaluate each rod's response with a separate prediction, by centering the network's field of view sequentially on each nanostructure, assuming that short-range interactions inside the predicted area are the most significant source of the local field's perturbation. This is depicted in figure~\ref{fig:neural_network_model}b, our goal is to perform calculations of arbitrarily large, disordered metasurfaces. 
	In the following, we refer to this procedure as ``dressed polarizability substitution''.
	
	This approach should provide a good approximation in the case when optical interactions are mostly short-range, and can be well approximated by taking interactions into account only within the local field of view of the artificial neural network.
	To qualitatively demonstrate the validity of this local neighborhood approximation, we provide a first comparison with a full simulation in figure~\ref{fig:neural_network_model}c, showing excellent agreement. 
	A more detailed analysis and benchmarks will be given later in the following.
	
	In general, the approach is not restricted to a specific type of nanostructure. 
	Within the limits imposed by the discretization of the top-view image, it can handle nano-scatterers of any possible geometry and size -- under the validity conditions of the dressed polarizability approximation, which is discussed above.
	Also, while the height of the structures needs to be constant with the chosen 2D CNN, variable height structures would be possible for instance through a 3D CNN.\cite{wiechaDeepLearningMeets2020}
	For the purpose of the demonstrations in this study, we use random meta-surfaces composed of identical gold nanorods of $70\times 30\times 30\,$nm$^3$, covering $7\times 3$ pixels in the top-view images. 
	All rods are aligned with their long axis along $X$.
	Using the same geometry for every nano-structure in the random assemblies implies that a modification of the optical response is purely due to optical coupling.
	To furthermore demonstrate that the dressed polarizability concept is capable to describe inhomogeneous environments, we deposit the nanorods on a glass substrate ($n_{\text{subst}}=1.45$) surrounded by air ($n_{\text{env}}=1$).
	For a first test we spectrally limit the data-set to the resonance wavelength of a single rod at $\lambda_0=635\,$nm.

	\paragraph*{Artificial neural network architecture}
	
	We use a 2D convolutional neural network, following the U-Net design.\cite{ronnebergerUNetConvolutionalNetworks2015} 
	The network is composed of several consecutive residual blocks.\cite{heDeepResidualLearning2015, szegedyInceptionv4InceptionResNetImpact2016}
	The input of the network is composed of two channels. The first channel describes the spatial distribution of the nano-scatterers in a 2D projection onto the substrate plane (``top view''). In our demonstration we discretize the scatterers and their positions on a grid of $10\,$nm stepsize.
	The wavelength of the illumination is provided through the second input layer, as indicated on the left of figure~\ref{fig:neural_network_model}a.
	The output layer of the network has the same 2D spatial dimension as the input and consists of 18 channels, 9 for the real and 9 for the imaginary part of the components of the $3\times 3$ dressed polarizability tensors (see right of figure~\ref{fig:neural_network_model}a).
	For training of the network, the components of the polarizability tensor of a nanostructure are therefore copied to every pixel covered by the geometry in the top-view image. 
	From the predicted 2D maps of the trained network, we then take the mean value of all pixels corresponding to each single structure.
	Once training is finished, a prediction takes around $2\,$ms on the same GPU used for training, whereas the simulation takes between $5$ and $60$ seconds, depending on the density of gold nanorods.
	
	All details on the neural network and training hyperparameters can be found in the \manualref{supporting information (SI) section A}.
	A description of the output format and its processing can be found in the \manualref{SI section B}, the training convergence is discussed in the \manualref{SI section C}.

	\paragraph*{Training data generation}
	
	For the training of the neural network we generate a dataset of 65,000 random arrangements of which 50,000 samples consist of ``fully filled'' areas with an average number of $55 \pm 6$ nanorods. 
	The remaining 15,000 samples consist of truncated structures, where the nanorods are removed on one or several sides of the area to simulate partial filling.
	Details on the geometry generation procedure, pre-processing as well as statistics on the numbers of rods in the training data can be found in the \manualref{SI section B}.
	Data generation takes about three weeks on a PC with an 8th generation intel i7 processor, $64\,$GB RAM and an Nvidia RTX 2060 SUPER graphics processing unit (GPU), using a CUDA-accelerated solver for GPU based extraction of the dressed polarizabilities through full-field simulations.\cite{wiechaPyGDMNewFunctionalities2022}

	\begin{figure}[t]
		\centering
		\includegraphics[width=.9\columnwidth]{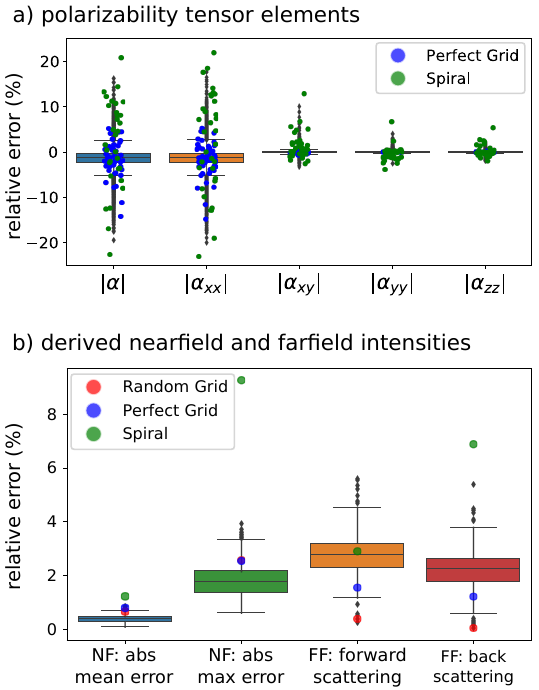}
		\caption{
			Statistics of the prediction accuracy.
			a) box plots showing statistics of the prediction error for the polarizability tensor components. The statistics are based on a test-set of 1000 samples containing around 45000 nanorods.
			The errors are given relative to the Frobenius norm of each polarizability tensor.
			The blue, respectively green plots show the errors of the individual rods of the examples shown in figure~\ref{fig:prediction_examples_fixed_wl}b, respectively~\ref{fig:prediction_examples_fixed_wl}c.
			b) box plots of the prediction accuracy of derived observables for the same 1000 test geometries. 
			From left to right are shown average and peak error of the near-field intensity in a plane $150\,$nm above the substrate, and the error of the far-field intensity in the upper (forward scattering towards the air side) as well as in the lower hemisphere (back scattering into the substrate). All errors are relative to the simulated results.
			The colored round markers correspond to the accuracies of the according examples in figure~\ref{fig:prediction_examples_fixed_wl}.
		}
		\label{fig:statistics}
	\end{figure}

	\section*{Results and Discussion}

	\paragraph*{Selected examples}
	
	In figure~\ref{fig:prediction_examples_fixed_wl} we show three specific examples of predicted structures and compare them to full field simulations.
	Figure~\ref{fig:prediction_examples_fixed_wl}a shows a structure generated in the same manner as the training and validation/test data.
	In the figures~\ref{fig:prediction_examples_fixed_wl}b and~\ref{fig:prediction_examples_fixed_wl}c we show respectively, a perfectly ordered array of $7\times 7$ rods, and a spiral of gold nanorods. 
	The latter corresponds to an ordered geometry without periodicity.
	The left two columns compare the predicted component $\alpha_{xx}$ (left-most column) with the respective values extracted from simulations (second left panel).
	The two center columns compare in the same way the $\alpha_{xy}$ component. The latter is zero in the isolated gold-rod, hence this component emerges as a result of optical interactions.
	On the right, we compare the near-field intensity in a plane $150\,$nm above the substrate, for a normal incidence plane wave arriving from inside the substrate with left circular polarization (LCP). The wavelength $\lambda_0=635\,$nm corresponds to the isolated rod's resonance condition.
	We find an excellent agreement with simulations in all examples which is remarkable especially because ordered structures like the examples in Fig.~\ref{fig:prediction_examples_fixed_wl}b and~\ref{fig:prediction_examples_fixed_wl}c were not explicitly present in the training data.
	2D maps of the network errors relative to the full simulations for all examples are given in the \manualref{supporting information section D}.

	\begin{figure*}[t]
		\centering
		\includegraphics[width=\linewidth]{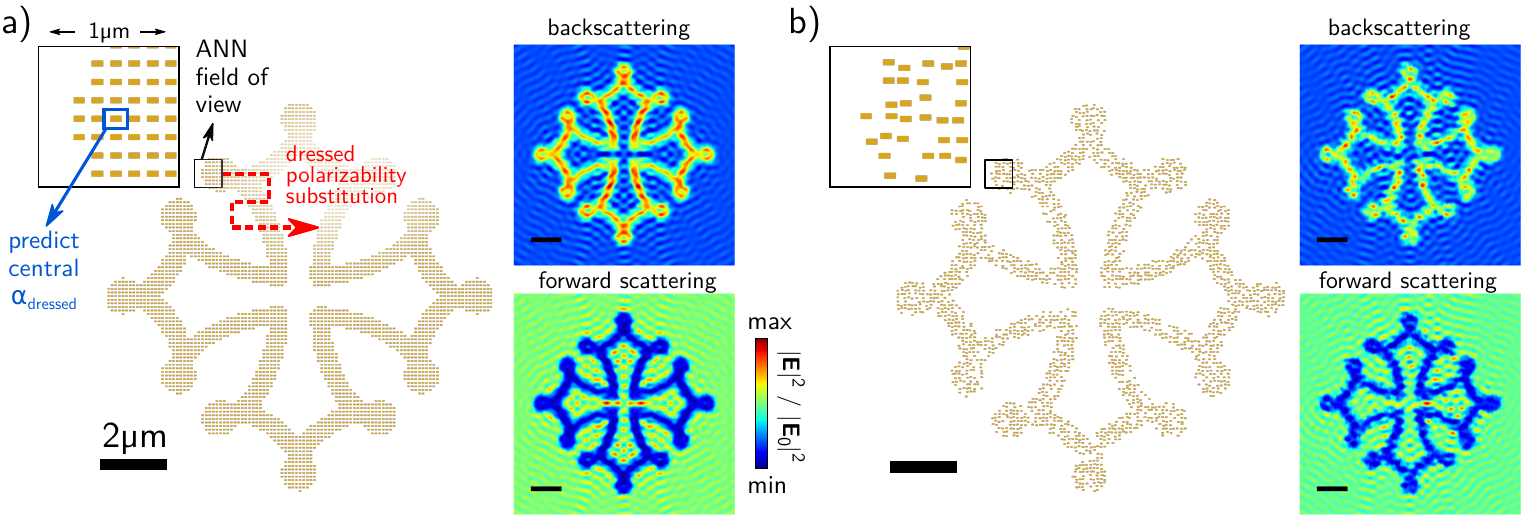}
		\caption{
			a) Example of a calculation based on the local neighborhood approximation of a large non-periodic rod-assembly, the rods forming the structure are distributed in an equidistant manner. The structure consists of $3016$ gold rods.
			The ``dressed polarizability substitution'' procedure, based on the short-range interaction assumption, is depicted in the sketch of the geometry.
			b) same as (a) but with a random displacement of the rods, overlapping rods are removed. The disordered structure consists of $2262$ gold rods.
			The panels on the right of (a-b) show field intensity maps $150\,$nm above  the substrate for an illumination from above (backscattering), respectively below (forward scattering).
			If not indicated otherwise, scale bars correspond to $2\,$\textmu m, the zoom insets show areas of $1\times 1\,$\textmu m$^2$, corresponding to the neural network input area.
			The nanorods in all examples are laying on a glass substrate ($n_{\text{subst}}=1.45$). The illumination wavelength is at the resonance  $\lambda_0=635\,$nm, with linear polarization along the rod axis.
		}
		\label{fig:local_approx}
	\end{figure*}

	\paragraph*{Statistical evaluation}
	
	We also perform a systematic benchmark of the prediction accuracy on a test set of 1,000 geometries, which are generated the same way as the training data (see example (a) ``random'' in figure~\ref{fig:prediction_examples_fixed_wl}). 
	The test set contains in total around 45,000 nanorods.
	The statistics of the predictions on those nanorods are depicted in figure~\ref{fig:statistics}a for the individual tensor components. 
	The interquartile range spans over errors smaller than $3\,$\%.
	Since the nanorods are oriented along $X$, the tensor component $\alpha_{xx}$ has the largest magnitude (c.f. Fig.~\ref{fig:polarization_sketch}d), and it is also the component with the highest uncertainty. 
	We found that a few outliers show errors of up to around $15\,$\% (black diamond markers), which affects around 5\% of the gold rods.
	In our configuration, the off diagonal elements of the dressed polarizabilities have considerably smaller magnitude (at least one order of magnitude smaller), and also their prediction errors relative to the full tensor norm are accordingly smaller. 
	We do not show the off-diagonal terms $\alpha_{{xz}}$ and $\alpha_{{yz}}$, which 
	are in general very weak for an assembly of structures lying in the $XY$ plane. 
	They have similarly small errors as the $\alpha_{zz}$ component.
	Figure~\ref{fig:statistics}b shows the accuracy of derived field intensities in the near- and far-field under plane wave illumination from within the substrate.
	These are obtained via re-propagation of the dipole moments induced by a normally incident, $X$-polarized plane wave (see equation~(\ref{eq:repropa_E})). 
	The near-field is calculated at a distance of $150\,$nm above the substrate, where we find very low average errors smaller than 1\%. 
	Taking the peak of the nearfield intensity error in each test sample gives errors of the order of 2\%. 
	The far-field is integrated over a sphere around the ensemble of nanorods and gives on average an accuracy of around 2.5\% for forward scattering (into the air environment), respectively 2\% for backward scattering (into the substrate).
	The statistics of the derived fields also has less outliers, of the order of 1\% or less.
	The errors of the derived near-field intensities are generally smaller than the error on the underlying predictions of dressed polarizabilities.
	We attribute this to an averaging effect in the dense ensemble of multiple scatterers. Due to their varying signs, the errors of the individual dressed polarizabilities partially cancel out in the near-field.
	In the far-field calculations on the other hand, the errors are commensurable with the peak near-field error, which we attribute to the fact that the strongest few dipoles will dominate far-field scattering. 
	Therefore, the error is again determined by the prediction accuracy of the individual polarizability tensors.
	
	To categorize the quantitative fidelity of the examples in figure~\ref{fig:prediction_examples_fixed_wl}, we compare the accuracies of these examples to the statistical evaluation.
	The red, blue and green markers in figure~\ref{fig:statistics} indicate, respectively, the errors of the random grid (Fig.~\ref{fig:prediction_examples_fixed_wl}a) of the ordered grid (Fig.~\ref{fig:prediction_examples_fixed_wl}b) and of the spiral structure (Fig.~\ref{fig:prediction_examples_fixed_wl}c).
	We find that the two ordered examples have a larger error margin compared to the random structure of same geometric appearance as used in training. 
	We believe that this is due to the absence of such ordered geometries in the training data, but the good agreement indicates that the ANN manages to generalize to cases with a different structural order and different densities of nanorods than used for training.

	\subsection*{Local neighborhood approximation for very large aperiodic metasurfaces}
	
	\begin{figure}[t]
		\centering
		\includegraphics[width=\linewidth]{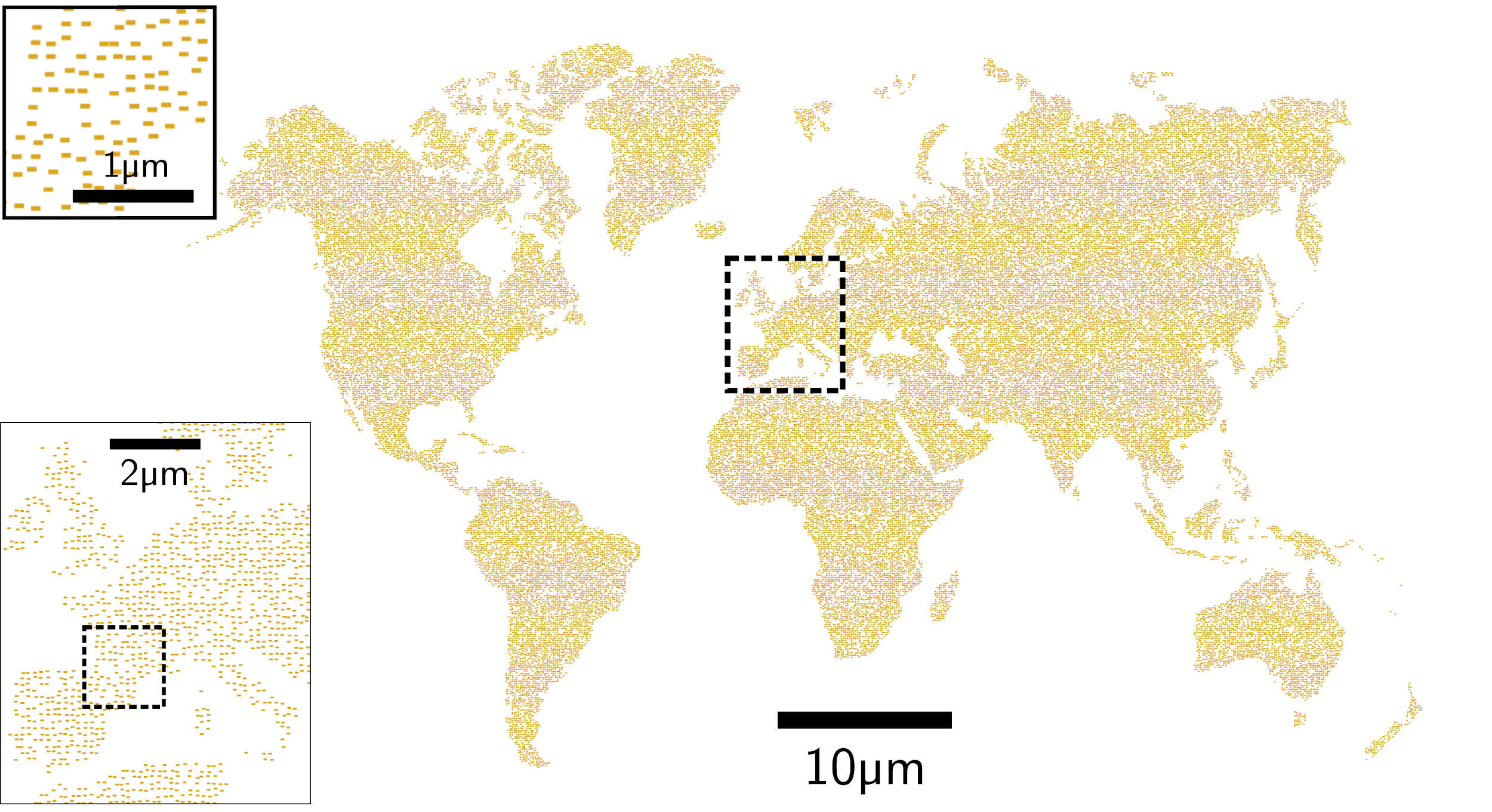}
		\vspace{0.5cm}
		
		\includegraphics[width=\linewidth]{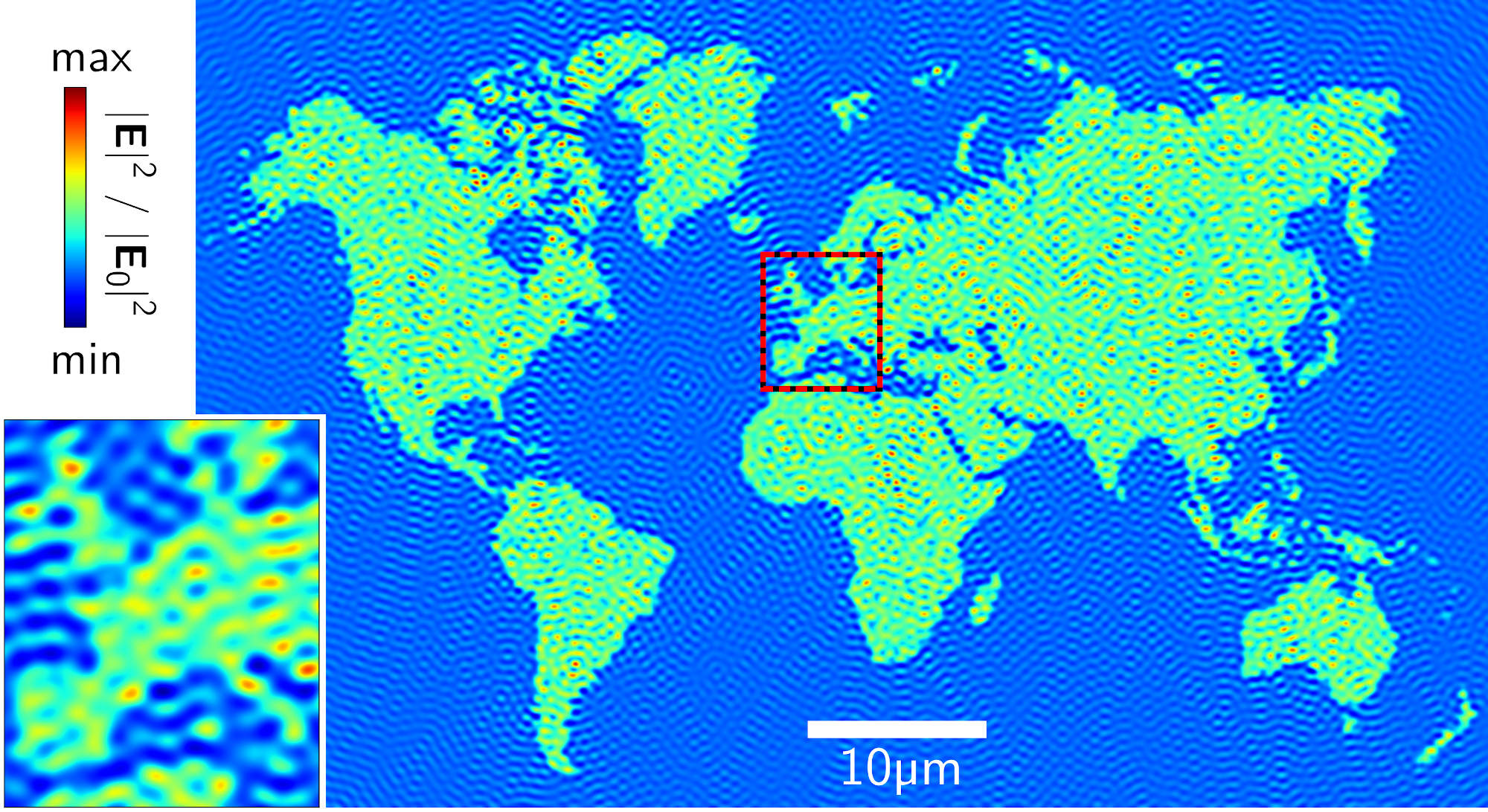}
		\caption{
			Local neighborhood approximation of a very large dis-ordered metasurface. 
			Top: Very large planar geometry formed by around 55,000 disordered gold nanorods lying on a glass substrate. The metastructure covers a surface of around $80\times 50\,$\textmu m$^2$.
			Bottom: Electric field intensity map at $250\,$nm height above the plane of gold rods, illuminated by a linearly polarized plane wave of $\lambda_0=635\,$nm, incident from the top.
			Insets on the left show zooms of the indicated central region, expanding over an area of $6.5\times 8\,$\textmu m$^2$.
			The inset in the  top left of the upper panel shows a further zoom into an area of $1.7\times 1.7\,$\textmu m$^2$ (area highlighed by a dashed frame in the bottom left inset of the upper panel). Based on the image ``world map blank'', wikimedia (2009), open access.
		}
		\label{fig:local_approx_very_large}
	\end{figure}

	Having verified the accuracy of the dressed polarizability predictor, our goal is now to use the network model in a sequential procedure for each individual element independently, to approximate the optical response of aperiodic plasmonic metasurfaces of essentially arbitrarily large size. 
	As mentioned in the beginning, to this end we apply a second approximation, which we call the ``\textit{local approximation}'', similar to what has been recently proposed at the level of a metasurface unit cell\cite{anDeepConvolutionalNeural2021}. 
	Therein we assume that optical interactions are predominantly short-range, hence that the dressed polarizability of a nano-structure is mainly defined by its close vicinity. 
	In other words, we presume that nanostructures at larger distances have little to no impact on the local optical field. 
	This assumption is undoubtedly false in case of a wave propagating in the plane of the nanostructures. On the other hand, for the normal incidence considered here, our hypothesis is likely to be justified.
	In prior tests we found that an interaction range of $\pm 500\,$nm for a wavelength of $\lambda_0=635\,$nm leads to good results with average accuracies around $2$-$3$\%.
	Thus our choice of an input area of $1\times 1\,$\textmu m$^2$. 
	An analysis of the local approximation accuracy as function of the interaction range is given in the \manualref{SI, section E}, where we also provide an error-map for the comparison of full-structure simulation and network-based sequential evaluation, given in figure~\ref{fig:neural_network_model}c.
	
	In the local approximation, we extract the dressed polarizability of each nanostructure independently throughout the metasurface. 
	To this end we position every nanostructure of the ensemble in the center of the ANN's input area. We remove all structures outside the network's field of view and feed the reduced area into the neural network for prediction.
	From the resulting dressed polarizabilities of the local scene, we discard all values except for the center nanostructure. 
	Thus we perform one prediction for each nanostructure in the large aperiodic metasurface.
	The scheme is illustrated in figure~\ref{fig:local_approx}a (c.f. also figure~\ref{fig:neural_network_model}b).
	Once all dressed polarizabilities are obtained, we reconstruct the full metasurface by attributing the respective $\alpha_{\text{dressed}}$ to the nanorods' absolute positions in the ensemble.
	Now we can illuminate the polarizabilities with a plane wave to obtain the effective electric dipole moments, which in turn can be repropagated to calculate near- and far-fields (via equations~(\ref{eq:repropa_E}) and~(\ref{eq:repropa_H})).
	Figure~\ref{fig:neural_network_model}b depicts a comparison of the ANN based dressed polarizability substitution method with a full field simulation, where an excellent agreement between the methods is observed. 
	
	The local approximation in combination with the accelerated polarizability calculation allows now to simulate assemblies of nano-structures (here gold nanorods) of essentially arbitrary size.
	The simulation cost scales linearly with the number of scatterers, the estimation of each scatterer's dressed polarizability takes around $2\,$ms on our hardware (see above). 
	The linear scaling might still appear unfavorable, but compared to many other techniques like the GDM\cite{wiechaPyGDMPythonToolkit2018}, discontinuous Galerkin\cite{fezouiConvergenceStabilityDiscontinuous2005}, boundary elements\cite{hohenesterMNPBEMMatlabToolbox2012} or finite elements\cite{gallinetNumericalMethodsNanophotonics2015}, a linear scaling is actually very attractive.
	As a first demonstration, we show in figure~\ref{fig:local_approx}a a large planar structure, composed of 3,016 gold nanorods. The structure has an extension of around $12\times 12\,$\textmu m$^2$, filled areas are occupied with periodic gold rods of $70\times 30\times 30\,$nm$^3$ size, placed with a center-to-center distance of $120\,$nm.
	We calculate the nearfield intensity in a parallel plane, $150\,$nm above the substrate under $X$-polarized plane wave illumination, incident either from the top (``backscattering''), or from below (``forward scattering'').
	We find that, as expected, the neural network based local approximation finds symmetric field intensity maps for the pattern with periodic nanorod filling. 
	In a second simulation shown in figure~\ref{fig:local_approx}b, we disturb the rod positions randomly with shift steps of $10\,$nm or $20\,$nm in random directions, while removing touching and overlapping rods. 
	The resulting geometry is shown on the left of Fig.~\ref{fig:local_approx}b, the according simulated near-field intensity maps on the right.
	Introducing randomness breaks the regular character of the nearfield distribution.
	
	As a visual demonstration of the capability to simulate very large, totally non-periodic metasurfaces, we show in figure~\ref{fig:local_approx_very_large} the near-field intensity after plane wave illumination close to a very large assembly of around 55,000 gold nanorods. The structure extends over an area of around $80\times 50\,$\textmu m$^2$ and was generated with the same randomization procedure as figure~\ref{fig:local_approx}b, the rods are hence very dense and near-field interactions play an important role on a local scale.
	The prediction of the dressed polarizabilities takes around 2 minutes, whereas a calculation with the same local approximation but using conventional simulations would require a few weeks of calculation time on the same hardware.

	\begin{figure}[t]
		\centering
		\includegraphics[width=\columnwidth]{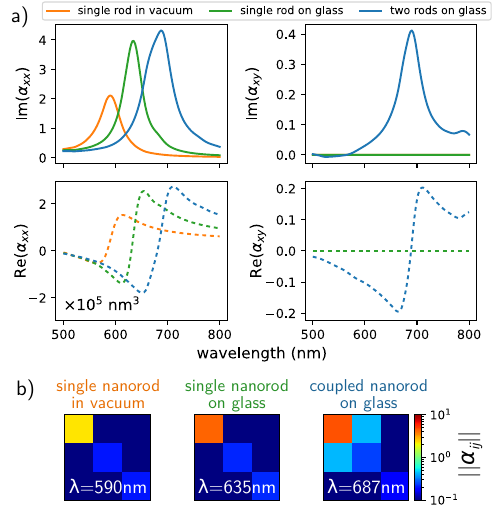}
		\caption{
			(a) spectra of the dressed polarizability components $\alpha_{xx}$ (left column) and $\alpha_{xy}$ (right column) of an individual gold rod in vacuum (orange lines), on a glass substrate (green lines) and on glass next to a second nanorod of the same dimensions (blue lines). Top: imaginary parts (solid lines) and bottom: real parts (dashed lines).
			(b) Norm of the dressed polarizability components at the resonance wavelength of $\text{Im}(\alpha_{xx})$.
		}
		\label{fig:polarization_sketch_spectral}
	\end{figure}

	\begin{figure*}[t]
			\centering
			\includegraphics[width=0.99\linewidth]{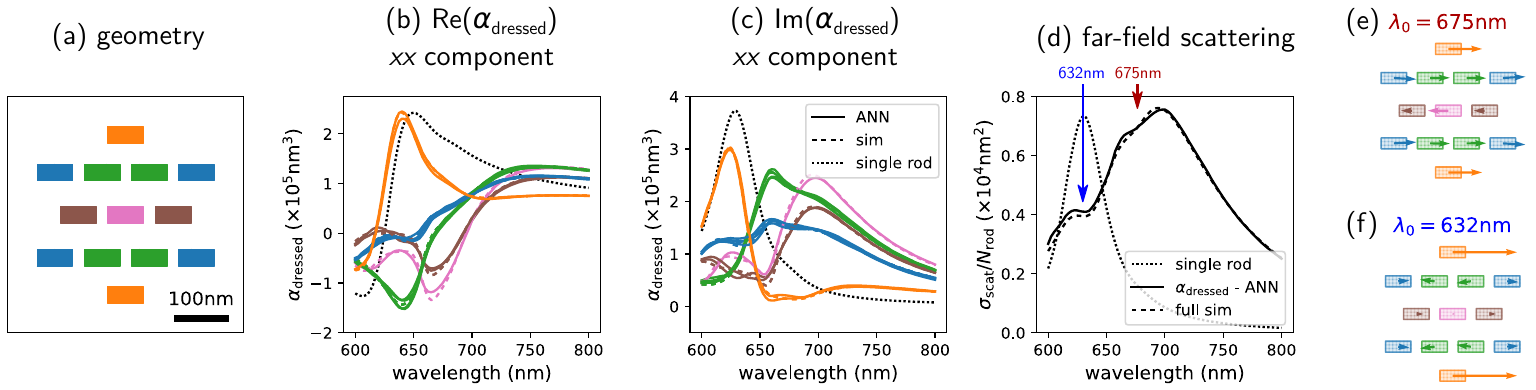}
			\caption{Spectral predictions of coupling effects in a plasmonic oligomer structure.
					a) Geometry of the studied oligomer composed of 13 gold nano-rods. The color-code regroups elements for which symmetry imposes identical $xx$ components of the dressed polarizability tensor.
					b) real part of the $xx$ components of all gold rods. Dashed lines correspond to simulations, solid lines to the prediction of the neural network model. The color-code is identical to a).
					c) same as b) but showing the imaginary part.
					d) far-field scattering cross section, calculated for $X$-polarized plane wave illumination at normal incidence. simulation (dashed line) and network prediction (solid line) are in good agreement.
					The dotted black lines in (b-d) correspond to the simulation of a single, isolated gold rod.
					(e-f) Sketch of the predicted effective dipole moment vectors in each gold rod (top view projection). The two panels correspond to the two small dips in the scattering spectrum. (e) $\lambda_0=675\,$nm and (f) $\lambda_0=632\,$nm. The wavelengths are highlighted in (d) by small arrows.
				}
			\label{fig:prediction_spectral_oligomer}
	\end{figure*}

	\subsection*{Spectral predictions}
	
	In the last section we show how the neural network model can be used in a wavelength dependent prediction scenario.
	Spectrally resolved predictions are particularly interesting as the local environment can have a strong impact on the resonance wavelength of a nanostructure, modifying its width or the spectral evolution of the induced phase shift. 
	This is depicted in Fig.~\ref{fig:polarization_sketch_spectral}a, where a single gold nanorod in vacuum (orange lines) is compared to the same nanorod on a glass substrate (green lines) as well as to a gold rod in close vicinity to a second nanorod (blue lines).
	Figure~\ref{fig:polarization_sketch_spectral}b shows the norm of the dressed polarizability tensor elements at the respective resonance wavelength.
	Coupling between plasmonic particles can furthermore induce effects such as mode hybridization, which can be identified conveniently in a spectral analysis.
	
	To extend the deep learning model to spectral predictions, we pass the wavelength into our ANN via a separate input layer (see figure~\ref{fig:neural_network_model}a, symbol ``$\lambda$'').
	We train this neural network on a second data set, using again the fixed geometry of $70\times 30\times 30\,$nm$^3$ gold nano-rods for every constituent. The spectral training dataset consists of 150,000 randomized simulations, using the same gold rod densities and truncation procedure as for the fixed wavelength dataset, however now with slightly smaller windows of $600\times 600$\,nm$^2$ to accelerate data generation (see also \manualref{SI, section~B}).
	We calculate dressed polarizabilities at wavelengths between $\lambda_0=600\,$nm and $\lambda_0=800\,$nm. The wavelengths are randomly chosen on discrete steps of $10$\,nm. 
	Each random geometry is thereby evaluated at only a single, random wavelength, thus the network has to learn implicitly the spectral correlations in the data, like the continuity of the spectra.
	In comparison to the fixed-wavelength network we find a similar prediction accuracy. A detailed analysis with more examples and statistics can be found in \manualref{SI, sections~F and~G}.

	To benchmark the capacity of the model for spectral predictions, we now use the predictor to reproduce effects in coupled systems of gold-nanostructure oligomers, reported in literature.\cite{zhangCoherentFanoResonances2013}
	Figure~\ref{fig:prediction_spectral_oligomer}a shows the geometry of the investigated oligomer, which is composed of 13 gold nano-rods. 
	Figure~\ref{fig:prediction_spectral_oligomer}b and~\ref{fig:prediction_spectral_oligomer}c show the real, respectively imaginary parts of the $xx$ component of the respective dressed polarizability tensors. Due to symmetry, several rods must have identical $xx$ components, which are indicated by the color-code.
	
	We find that the predicted dressed polarizability spectra are in very good agreement with the simulated values. The symmetry properties are well reproduced over the entire spectral range, even though the deviation from an isolated rod's response (dotted black lines) is considerable.
	We note that the quality of the prediction is in particular remarkable since the training data consists only of randomized structures, whereas here we test the model on an ordered structure of high symmetry.
	We finally use the dressed polarizabilities to calculate the far-field scattering spectrum of the structure under $X$-polarized normally incident plane wave illumination, which is shown in fig.~\ref{fig:prediction_spectral_oligomer}d. 
	We observe two dips in the scattering, at $\lambda_0 = 632$\,nm and  $\lambda_0 = 675$\,nm (indicated by a blue and red arrow, respectively), which are a result of opposing contributions in the local field, created due to coupling between the gold rods. 
	Figures~\ref{fig:prediction_spectral_oligomer}e and~\ref{fig:prediction_spectral_oligomer}f show the dipole moments at those two wavelengths, reproducing well the response previously reported in literature.\cite{zhangCoherentFanoResonances2013}
	The effect is just shifted to shorter wavelengths and is generally weaker, due to the geometry of the elements, different than in the literature example, where nano-discs were used.
	
	In conclusion, we demonstrated with a simple example that the dressed polarizability predictor network can be effectively trained on data with varying wavelengths. This paves the way to a multitude of possible applications, for instance homogenization of non-periodic plasmonic metasurfaces\cite{repanArtificialNeuralNetworks2021} for which we provide a simple example in the \manualref{supporting information section H}.
	The method could be useful also for the inverse design of dense, non-periodic plasmonic metasurfaces, to take into account optical interactions between meta-atoms.

	\section*{Conclusions}
	
	In conclusion, we demonstrated a deep learning based technique to predict the optical response of large, non-ordered and dense assemblies of optically interacting plasmonic nanostructures using dressed polarizabilities. 
	The method is capable to reduce the high computational cost of calculating the complex interactions in large, aperiodic arrangements of nanostructures by five orders of magnitude.
	We showed that the model can be applied efficiently to describe the spectral response of dispersive materials (in our case gold).
	In the limits of a local approximation, neglecting long-range optical interactions, we furthermore demonstrated that the method can be used to describe very efficiently the optical response of planar assemblies of nano-structures of virtually any size.
	
	Our approach can be straightforwardly generalized to nanostructures of arbitrary shapes or multiple materials, the first simply via the top-view discretization, the latter by using the spatial distribution of complex permittivity as input for the neural network.
	Also the refractive index of the environment, a substrate, a superstrate, etc. could be included in the model as an additional input parameter.
	It might also be generalized to oblique angles of incidence using non-local polarizability concepts\cite{bertrandGlobalPolarizabilityMatrix2019} or further approximations.\cite{patouxPolarizabilitiesComplexIndividual2020}
	The method is useful for the homogenization of totally disordered structures or aperiodic metasurfaces, but also of ordered geometries with large-scale or no periodicity.\cite{smithHomogenizationMetamaterialsField2006, braunEffectiveOpticalProperties2006} 
	The high computational efficiency renders the approach attractive for application in iterative optimization schemes, for instance to tailor the effective refractive index of a synthetic nanostructured layer, or for micron-scale wavefront-shaping, capable to take into account local coupling and optical interaction effects inside the nano-structured medium.
	We believe that our concept is of highest interest for a broad audience in the field of photonics and nano-optics, and in particular for the design and description of metasurfaces, where rigorous calculations of optical interactions in complex structures remain an important challenge.

	\subsection*{Associated  Content}
	\noindent
	Supporting information. A pdf providing details on the neural network architecture, the training procedure, as well as additional results and further analysis.
	This material is available free of charge via the internet at http://pubs.acs.org. Brief descriptions in non-sentence format listing the contents of the files supplied as Supporting Information.

	\subsection*{Funding}
	\noindent
	This work was supported by the German Research Foundation (DFG) through a research fellowship (WI 5261/1-1) and by the Toulouse HPC CALMIP (grant p20010). 
	OM acknowledges support through EPSRC grant EP/M009122/1.

	\subsection*{Notes}
	\noindent
	The authors declare no competing financial interest.

	\subsection*{Acknowledgments}
	\noindent
	We thank Caroline Bonafos, Aurélien Cuche and Adelin Patoux for fruitful discussions.
	We thank the NVIDIA Corporation for the donation of a Quadro P6000 GPU used for this research
	All data supporting this study are openly available from the University of Southampton repository (DOI: \href{https://doi.org/10.5258/SOTON/D2063}{10.5258/SOTON/D2063}).

	\def\bibsection{\section*{References}}
	\bibliography{2021_majorel_dressed_polarizability_predictor.bbl}

\end{document}